\documentclass[prc,preprint,showpacs,preprintnumbers,amsmath,amssymb]{revtex4}

\usepackage[mathscr]{eucal}
\usepackage{graphicx}
\def\slr#1{\setbox0=\hbox{$#1$}           
   \dimen0=\wd0                                 
   \setbox1=\hbox{/} \dimen1=\wd1               
   \ifdim\dimen0>\dimen1                        
      \rlap{\hbox to \dimen0{\hfil/\hfil}}      
      #1                                        
   \else                                        
      \rlap{\hbox to \dimen1{\hfil$#1$\hfil}}   
      /                                         
   \fi}

\def\ksq{k^2}

\def\mytint#1{\!\int\!\!\frac{d^3\!{#1}}{(2\pi)^3}\,}
\def\gev#1{ GeV${}^{#1}$}
\def\be{\begin{eqnarray}}
\def\ee{\end{eqnarray}}

\renewcommand{\theequation}%
    {\arabic{section}.\arabic{equation}}
\makeatletter \@addtoreset{equation}{section} \makeatother

\begin{document}

\preprint{BCCNT: 04/03/322}

\title{Quark Propagation in the Quark-Gluon Plasma}

\author{Xiangdong Li}
\affiliation{%
Department of Computer System Technology\\
New York City College of Technology of the City University of New
York\\
Brooklyn, New York 11201 }%

\author{Hu Li}
\author{C. M. Shakin}
\email[email:]{casbc@cunyvm.cuny.edu}
\author{Qing Sun}

\affiliation{%
Department of Physics and Center for Nuclear Theory\\
Brooklyn College of the City University of New York\\
Brooklyn, New York 11210
}%

\date{March, 2004}

\begin{abstract}
It has recently been suggested that the quark-gluon plasma formed
in heavy-ion collisions behaves as a nearly ideal fluid. That
behavior may be understood if the quark and antiquark mean-free-
paths are very small in the system, leading to a ``sticky
molasses" description of the plasma, as advocated by the Stony
Brook group. This behavior may be traced to the fact that there
are relatively low-energy $q\overline{q}$ resonance states in the
plasma leading to very large scattering lengths for the quarks.
These resonances have been found in lattice simulation of QCD
using the maximum entropy method (MEM). We have used a chiral
quark model, which provides a simple representation of effects due
to instanton dynamics, to study the resonances obtained using the
MEM scheme. In the present work we use our model to study the
optical potential of a quark in the quark-gluon plasma and
calculate the quark mean-free-path. Our results represent a
specific example of the dynamics of the plasma as described by the
Stony Brook group.
\end{abstract}


\maketitle

\section{INTRODUCTION}
The description of the quark-gluon plasma in terms of
hydrodynamics has been advocated by the Stony Brook group
[\,1-3\,]. That description appears to be in accord with the
experimental data. In such a description the motion of the quarks
is characterized by an extremely short mean-free-path. The origin
of that behavior is thought to be due to the relatively low-energy
resonances in the $q\overline{q}$ system leading to very large
scattering lengths. These resonances have been found in lattice
studies of QCD which make use of the maximum entropy method (MEM)
[\,4-9\,]. Similar resonances are found in the scalar,
pseudoscalar, vector and axial-vector $q\overline{q}$ channels
[\,10\,]. Recently, an extensive exploration of charmonium studies
in the confined and deconfined regions using lattice methods has
been reported in Ref. [11]. In that work results are given for the
dependence of the resonance excitation on the total momentum of
the $q\overline{q}$ pair. We have studied that dependence for
light quark systems in Ref.[12] and have found similar behavior to
that reported in Ref. [11]. (We will make use of the results
presented in Ref. [12] in the present work in which we calculate
the imaginary part of the optical potential and the mean-free-path
for a quark in the quark-gluon plasma.) We use a chiral model with
rather large momentum cutoff. That model is meant to provide an
approximate description of the instanton dynamics advocated by the
Stony Brook group [1-3]. Earlier work using our model may be found
in Refs. [13-15].

In our studies of meson spectra at $T=0$ and at $T<T_c$ we have
made use of the Nambu--Jona-Lasinio (NJL) model. The Lagrangian of
the generalized NJL model we have used in our studies is

\begin{flushleft}
\be \mathcal L&=&\overline{q}(i\slr\gamma-m^0)q+\frac{\overline
G_S}{2}\sum_{i=0}^8 [(\overline{q} \lambda^{i} q)^2+(\overline{q}
i \gamma_5 \lambda^{i} q)^2]\\\nonumber &-&\frac{\overline
G_V}{2}\sum_{i=0}^{8}[(\overline{q} \lambda^{i}\gamma_\mu
q)^2+(\overline{q} \lambda^{i}\gamma_5\gamma_\mu q)^2]\\\nonumber
&+&\frac{G_D}{2} \lbrace
\det[\overline{q}(1+\lambda_5)q]+\det[\overline{q}(1-\lambda_5)q]\rbrace
+\mathcal L_{conf} \ee
\end{flushleft}

Here, $m^0$ is a current quark mass matrix, $m^0=diag(m_u^0,
m_d^0, m_s^0)$. The $\lambda_i$ are the Gell-Mann (flavor)
matrices and $\lambda^0=\sqrt{2/3}\mathbf{1}$, with $\mathbf{1}$
being the unit matrix. The fourth term is the 't Hooft interaction
and $\mathcal L_{conf}$ represents the model of confinement used
in our studies of meson properties.

In the study of hadronic current correlators [13-15] it is
important to use a model which respects chiral symmetry when
$m^0=0$. Therefore, we make use of the Lagrangian of Eq. (1.1),
while neglecting the 't Hooft interaction and $\mathcal L_{conf}$.
In order to make contact with the results of lattice simulations
we use the model with the number of flavors $N_f=1$. Therefore,
the $\lambda^i$ matrices in Eq. (1.1) may be replaced by unity. We
then use \be \mathcal
L&=&\overline{q}(i\slr\gamma-m^0)q+\frac{G_S}{2}[(\overline{q}q)^2+(\overline{q}
i \gamma_5 q)^2]\\\nonumber &-&\frac{G_V}{2}[(\overline{q}
\gamma_\mu q)^2+(\overline{q}\gamma_5\gamma_\mu q)^2] \ee in order
to calculate the hadronic current correlation functions. Thus,
there are essentially three parameters to consider, $G_S$, $G_V$
and a Gaussian cutoff parameter $\alpha$, which restricts the
momentum integrals through a factor
$\exp[-\overrightarrow{k}^2/\alpha^2]$. As suggested by the Stony
Brook group, we consider the NJL model and the associated chiral
Lagrangian of Eq. (1.2) as providing a simplified representation
of the instanton dynamics important for the problems considered in
this work. Since the results obtained for the hadronic current
correlation functions are similar in the scalar, pseudoscalar,
vector and axial-vector channels, we carry out our calculations
for the scalar $q\overline{q}$ states and multiple our results for
the optical potential by 4. The parameters \emph{G} and $\alpha$
were fixed in our earlier studies [12]. We take \emph{G} =1.0
GeV$^{-2}$ and $\alpha$ = 4.4 GeV. These values provide good fits
[12] to the hadronic current correlation functions found in the
lattice studies [10]. In order to calculate the optical potential
for a quark we consider the quark moving in an antiquark
distribution characterized by a temperature-dependent occupation
factor $n(\overrightarrow{p_{1}})$ which depends upon the chemical
potential $\mu$. The energy of a quark is given by
$E(\overrightarrow{p})=[\overrightarrow{p}^{2}+m^{2}]^{1/2}$. We
follow the work of Shuryak [1], for example, and put $m$ = 1 GeV.
In Shuryak's work this mass is not the current quark mass, but is
called the ``chiral mass". (We would prefer to call the 1 GeV
mass, the ``thermal mass", however, the terminology used is not
important for this work.) The quark thermal mass is given in Ref.
[16], with $C_{F}=4/3$, as \be m^{2} =
\frac{1}{8}g^{2}C_{F}(T^{2}+\frac{\mu^{2}}{\pi^{2}}), \ee for the
case of a finite chemical potential. The thermal gluon mass is \be
m_{g}^{2} = \frac{1}{6}g^{2}T^{2}(C_{A}+\frac{1}{2}N_{f}). \ee
(The relation between thermal masses in QED and QCD is given on
p.146 of Ref. [\,16\,].) In studies of baryon matter, the chemical
potentials used are often about 300 MeV or less. However, once we
introduce a thermal mass of about 1 GeV, we need to determine the
chemical potential for the quarks. In this work we will consider a
chemical potential of about 1 GeV, although the calculations are
easily made for other values. Once we put $m$ = 1 GeV, the
chemical potential is the only parameter which is varied in our
study. The organization of our work is as follows. In Section II
we discuss the calculation of the imaginary part of the quark
optical potential and the quark mean-free-path. Section III
contains some further discussion and conclusions. (Appendixes A
and B provide a description of the calculation of correlators in
our model. That description is readily taken over to obtain a
representation of the $q\overline{q}$ scattering matrix. The
calculation of the vacuum polarization function
$J(P^{0},\overrightarrow{P}$) which appears in Section II is
described in the appendices.)

\section{calculation of the quark optical potential}

It is useful to contrast the calculation of the quark optical
potential with the calculation of the nucleon optical potential
that is to be used in the Dirac equation. The latter calculation
is discussed in detail in Ref. [17]. In that calculation of the
nucleon-nucleus potential one calculates the \emph{T}-matrix for
nucleon-nucleon scattering using the one-boson-exchange (OBE)
model. In that case the mesons of the OBE model undergo
\emph{t}-channel and \emph{u}-channel exchange between the
nucleons. The result is that the imaginary part of the optical
potential has a magnitude of about 10 MeV [17]. That in turn leads
to a mean-free-path of about 10 fm for a 500 MeV nucleon. Many
years ago, the relatively large value for the nucleon
mean-free-path lead to the characterization of the optical model
for nucleon-nucleus scattering as the ``cloudy crystal ball"
model. When we study quark propagation in the quark-gluon plasma
we may consider a similar calculation of the optical potential at
finite temperature. In the case of the quark-antiquark interaction
the \emph{T} matrix is dominated by \emph{s}-channel resonances of
the type found in the MEM studies. As we will see, the interaction
in this case is quite strong, leading to a small mean-free-path.
The resulting model is called the ``sticky molasses" model [1-3]
as opposed to the ``cloudy crystal ball" model used to describe
nucleon-nucleus scattering. We now consider the potential seen by
a quark of momentum $\overrightarrow{p_{2}}$ and average over the
quark spin $s_{2}$. (We will consider quarks of a single flavor,
since that was done in the MEM studies that we have used to fix
the parameters of our model.) In Ref. [17] the relativistic
optical potential was denoted as $\Sigma(\overrightarrow{p},s)$
and, for this work, we consider \be
\Sigma^{++}(\overrightarrow{p_{2}}) =
\frac{1}{2}\sum_{s_2}\overline{u}(\overrightarrow{p_{2}},s_{2})\Sigma(\overrightarrow{p_{2}},s_{2})u(\overrightarrow{p_{2}},s_{2}).
\ee

It is useful to introduce [14] \be U(\overrightarrow{p_{2}}) = N
\sqrt{\frac{m}{E(\overrightarrow{p_{2}})}}\Sigma^{++}(\overrightarrow{p_{2}})\sqrt{\frac{m}{E(\overrightarrow{p_{2}})}}.
\ee

Here the factor of \emph{N} = 4 takes into account the sum of the
interactions in the scalar, pseudoscalar, vector and axial-vector
channels which are taken to be equal for the purposes of this
work. The approximate equality of the interactions in these
channels may be seen in Ref. [10]. (Note that values of
$U(\overrightarrow{p})$ are given in Ref. [17] for the case of
nucleon-nucleus scattering.)

If $p_{1}$ is the momentum of the antiquark in the medium, we may
introduce the four-vector \be P^{\mu} = (p_{1} + p_{2})^{\mu}. \ee
Now \be P^{2} &=&  P_{0}^{2} - \overrightarrow{P}^{2} \\
 &=&(E(\overrightarrow{p_{1}})+E(\overrightarrow{p_{2}}))^{2} -
 (\overrightarrow{p_{1}}^{2}+\overrightarrow{p_{2}}^{2}+2p_{1}p_{2}\cos\theta).
\ee

Here, we take $\overrightarrow{p_{2}}$ along the $z$ axis. We
define \be t(\overrightarrow{p_{1}},\overrightarrow{p_{2}}) =
\frac{1}{\pi P^{2}}
\left[\frac{G}{1-GJ(\overrightarrow{p_{1}},\overrightarrow{p_{2}})}\right],
\ee where $J(\overrightarrow{p_{1}},\overrightarrow{p_{2}})$ is
the $q\overline{q}$ vacuum polarization function defined in
Appendix B. (We remark that we may also use the notation
$t(P^{2},p_{2})$ for the quantity defined in Eq. (2.6).)

It is also useful to introduce the occupation factor
$n(\overrightarrow{p_{1}})$: \be n(\overrightarrow{p_{1}}) =
\frac{1}{\exp\beta[E(\overrightarrow{p_{1}})-\mu]+1}, \ee with
$\beta=1/T$ and
$E(\overrightarrow{p_{1}})=\left[\overrightarrow{p_{1}}^{2}+m^{2}\right]^{1/2}$.
We recall \be \sum_{s_2} u(p_{2},s_{2})\overline{u}(p_{2},s_{2}) =
\left(\frac{\slr p_{2} + m}{2m}\right), \ee \be \sum_{s_1}
v(p_{1},s_{1})\overline{v}(p_{1},s_{1}) = \left(\frac{\slr p_{1} -
m}{2m}\right), \ee  and note that

\be \mbox{Tr}\left(\frac{\slr p_{2}+m}{2m}\right)\left(\frac{\slr
p_{1}-m}{2m}\right) =
\left[\frac{(E_{1}E_{2}-\overrightarrow{p_{1}}\cdot\overrightarrow{p_{2}})-m^{2}}{m^{2}}\right].
\ee

Thus, \be \mbox{Im}\Sigma^{++}(p_{2}) = -\frac{1}{2}\int
\frac{d\overrightarrow{p_{1}}}{(2\pi)^{3}}\frac{m}{E(\overrightarrow{p_{1}})}
\mbox{Im}\left[\frac{G}{1-GJ(\overrightarrow{p_{1}},\overrightarrow{p_{2}})}\right]
\\\nonumber
\times\left[\frac{E_{1}E_{2}-\overrightarrow{p_{1}}\cdot\overrightarrow{p_{2}}-m^{2}}{m^{2}}\right]n(\overrightarrow{p_{1}}),
\ee  and \be U(\overrightarrow{p_{2}}) =
-\frac{N}{2}\frac{m}{E(\overrightarrow{p_{2}})}\int\frac{d\overrightarrow{p_{1}}}{(2\pi)^{3}}\frac{m}{E(\overrightarrow{p_{1}})}\pi
P^{2}t(\overrightarrow{p_{1}},\overrightarrow{p_{2}})\\\nonumber
\times
\left[\frac{E_{1}E_{2}-\overrightarrow{p_{1}}\cdot\overrightarrow{p_{2}}-m^{2}}{m^{2}}\right]n(\overrightarrow{p_{1}}).
\ee

Here $E_{1}=E(\overrightarrow{p_{1}})$,
$E_{2}=E(\overrightarrow{p_{2}})$ and we have made use of Eqs.
(2.2) and (2.6). Values of $t(P^{2},p_{2})$ are shown in Fig. 1
for values of $|\overrightarrow{p_{2}}|$ ranging from 0.01 GeV to
0.31 GeV. In Fig. 2 we show the values of
$n(\overrightarrow{p_{1}})$ for the three values of $\mu$
considered here and in Fig. 3 we present values of
Im$U(\overrightarrow{p_{2}})$ for those values of $\mu$. In Fig. 4
we show the values for the mean-free-path \be \lambda =
\frac{|\overrightarrow{p_{2}}|}{m}\frac{1}{\mbox{Im}
U(\overrightarrow{p_{2}})}. \ee

\begin{figure}
\includegraphics[bb=0 0 280 235, angle=0, scale=1]{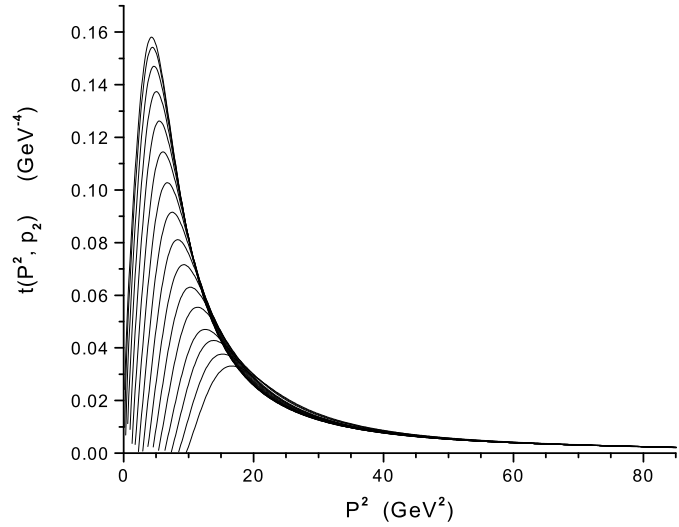}%
\caption{Values of $t(P^2, p_2)$ are shown for various values of
the quark momentum $|\vec{p}_2|$. Starting with the uppermost
curve, the $|\vec{p}_2|$ values in GeV units are 0.01, 0.03, 0.05,
0.07, 0.09, 0.11, 0.13, 0.15, 0.17, 0.19, 0.21, 0.23, 0.25, 0.27,
0.29 and 0.31. (For large $P^2$, we have $t(P^2, p_2)\simeq(1/\pi
P^2)G$.) Here $P^2=(p_1+p_2)^2$, where $p_1$ is the antiquark
momentum.}
\end{figure}

\begin{figure}
\includegraphics[bb=0 0 280 235, angle=0, scale=1]{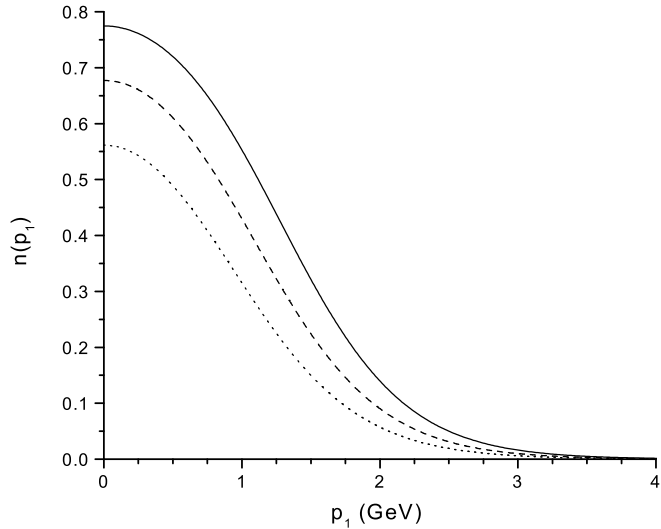}%
\caption{Values of $n(p_1)$ are shown for $\mu=1.1$\,GeV (dotted
curve), $\mu=1.3$\,GeV (dashed curve) and $\mu=1.5$\,GeV (solid
curve). Here $T=1.5\,T_c$ with $T_c=270$\,MeV.}
\end{figure}

\begin{figure}
\includegraphics[bb=0 0 280 235, angle=0, scale=1]{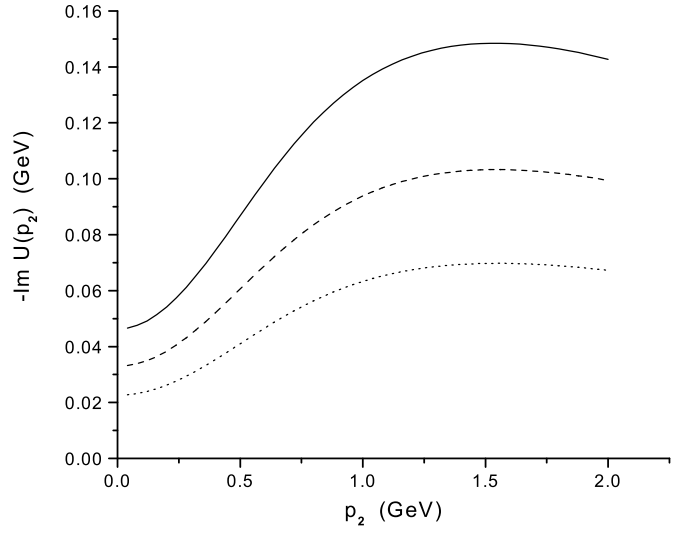}%
\caption{The imaginary part of the quark optical potential is
shown for $\mu=1.1$\,GeV (dotted curve), $\mu=1.3$\,GeV (dashed
curve) and $\mu=1.5$\,GeV (solid curve). (We recall that the
nucleon-nucleus imaginary optical potential is about 0.01\,GeV in
magnitude [17].)}
\end{figure}

\begin{figure}
\includegraphics[bb=0 0 280 235, angle=0, scale=1]{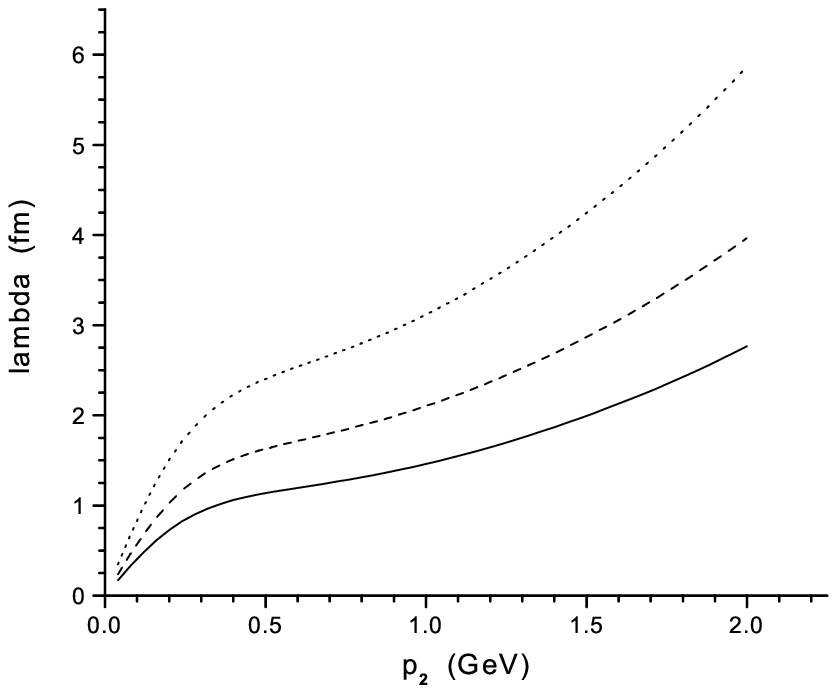}%
\caption{Values of $\lambda(p_2)$ are shown for $\mu=1.1$\,GeV
(dotted curve), $\mu=1.3$\,GeV (dashed curve) and $\mu=1.5$\,GeV
(solid curve).}
\end{figure}

\section{discussion}
Information is available concerning the \emph{baryon} chemical
potential. That chemical potential is parameterized in Ref. [18]
as \be \mu_{B} =
\frac{1270\,\mbox{MeV}}{(1+\frac{\sqrt{S_{NN}}}{4.3})}, \ee and
varies strongly with $\sqrt{S_{NN}}$, which is in GeV units in Eq.
(3.1). For $\sqrt{S_{NN}}$ = 200 GeV, we have $\mu_{B}$ = 26.7
MeV.

Of particular significance for our results is the choice of the
chemical potential for the quarks. We have taken $\mu \sim$ 1 GeV.
In the case of the quarks, the choice of $\mu \simeq$ 1 GeV leads
to small mean-free-paths consistent with the suggestion of Shuryak
that the resonances seen in the MEM analysis of the lattice
results are responsible for the small mean-free-paths of the
``sticky molasses" model.

We remark that in nuclear matter the baryon density is 0.17
fm$^{-3}$ or 0.51 quarks/fm$^{3}$ if we consider the nucleon to be
composed of three quarks. For the values of
$n(\overrightarrow{p_{1}})$ shown in Fig. 2 for the case $\mu$ =
1.3 GeV, we may calculate the density of antiquarks to be 5.91
fm$^{-3}$, so that the density of quarks and antiquarks is about
12 fm$^{-3}$ in our model. We remark that the energy density at
RHIC for $\sqrt{S_{NN}}$ = 200 GeV is 4.1 GeV/fm$^{3}$ [18], which
is about 26 times the energy density of nuclear matter, which is
approximately 0.16 GeV/fm$^{3}$.

In this work we have attempted to provide a quantitative analysis
of the suggestion [1] that the large $q\overline{q}$ resonant
scattering cross sections are responsible for the small quark
mean-free-paths, with the associated relevance of the hydrodynamic
description of the system that is created in high-energy
nucleus-nucleus collisions. We have considered the interaction in
the scalar, pseudoscalar, vector and axial-vector channels of the
$q\overline{q}$ system. It is possible that there are important
resonances of $qq$ character, as well as the $q\overline{q}$
resonances considered here. Such $qq$ states are depicted in Fig.
8a of Ref. [1]. It would be of interest to see if such states are
found in lattice studies using the MEM scheme.

\appendix
  \renewcommand{\theequation}{A\arabic{equation}}
  \setcounter{equation}{0}  
  \section{}  

For ease of reference, we present a discussion of our calculation
of hadronic current correlators taken from Ref.\,[\,15\,]. The
procedure we adopt is based upon the real-time finite-temperature
formalism, in which the imaginary part of the polarization
function may be calculated. Then, the real part of the function is
obtained using a dispersion relation. The result we need for this
work has been already given in the work of Kobes and Semenoff
[\,19\,]. (In Ref.\,[\,19\,] the quark momentum is $k^\mu$ and the
antiquark momentum is $k^\mu-P^\mu$. We will adopt that notation
in this section for ease of reference to the results presented in
Ref.\,[\,19\,].) With reference to Eq.\,(5.4) of Ref.\,[\,19\,],
we write the imaginary part of the scalar polarization function as
\be \mbox{Im}\,J_S(\textit{P}\,{}^2,
T)=\frac12N_c\beta_S\,\epsilon(\textit{P}\,{}^0)\mytint ke^{-\vec
k\,{}^2/\alpha^2}\left(\frac{2\pi}{2E_1(k)2E_2(k)}\right)\\\nonumber
\times\{[1-n_1(k)-n_2(k)]
\delta(\textit{P}\,{}^0-E_1(k)-E_2(k))\\\nonumber-[n_1(k)-n_2(k)]
\delta(\textit{P}\,{}^0+E_1(k)-E_2(k))\\\nonumber-[n_2(k)-n_1(k)]
\delta(\textit{P}\,{}^0-E_1(k)+E_2(k))\\\nonumber-[1-n_1(k)-n_2(k)]
\delta(\textit{P}\,{}^0+E_1(k)+E_2(k))\}\,.\ee Here,
$E_1(k)=[\,\vec k\,{}^2+m_1^2(T)\,]^{1/2}$. Relative to Eq.\,(5.4)
of Ref.\,[\,19\,], we have changed the sign, removed a factor of
$g^2$ and have included a statistical factor of $N_c$. In
addition, we have included a Gaussian regulator, $\exp[\,-\vec
k\,{}^2/\alpha^2\,]$. The value $\alpha=0.605$ GeV was used in our
applications of the NJL model in the calculation of meson
properties at $T=0$. We also note that \be
n_1(k)=\frac1{e^{\,\beta E_1(k)}+1}\,,\ee and \be
n_2(k)=\frac1{e^{\,\beta E_2(k)}+1}\,.\ee For the calculation of
the imaginary part of the polarization function, we may put
$\ksq=m_1^2(T)$ and $(k-P)^2=m_2^2(T)$, since in that calculation
the quark and antiquark are on-mass-shell. In Eq.\,(A1) the factor
$\beta_S$ arises from a trace involving Dirac matrices, such that
\be \beta_S&=&-\mbox{Tr}[\,(\slr k+m_1)(\slr k-\slr P+m_2)\,]\\
&=&2P^2-2(m_1+m_2)^2\,,\ee where $m_1$ and $m_2$ depend upon
temperature. In the frame where $\vec P=0$, and in the case
$m_1=m_2$, we have $\beta_S=2P_0^2(1-{4m^2}/{P_0^2})$. For the
scalar case, with $m_1=m_2$, we find \be \mbox{Im}\,J_S(P^2,
T)=\frac{N_cP_0^2}{8\pi}\left(1-\frac{4m^2(T)}{P_0^2}\right)^{3/2}
e^{-\vec k\,{}^2/\alpha^2}[\,1-2n_1(k)\,]\,,\ee where \be \vec
k\,{}^2=\frac{P_0^2}4-m^2(T)\,.\ee

For pseudoscalar mesons, we replace $\beta_S$ by
\be \beta_P&=&-\mbox{Tr}[\,i\gamma_5(\slr k+m_1)i\gamma_5(\slr k-\slr
P+m_2)\,]\\
&=&2P^2-2(m_1-m_2)^2\,,\ee which for $m_1=m_2$ is $\beta_P=2P_0^2$
in the frame where $\vec P=0$. We find, for the $\pi$ mesons, \be
\mbox{Im}\,J_P(P^2,T)=\frac{N_cP_0^2}{8\pi}\left(1-\frac{4m^2(T)}{P_0^2}\right)^{1/2}
e^{-\vec k\,{}^2/\alpha^2}[\,1-2n_1(k)\,]\,,\ee where $ \vec
k\,{}^2={P_0^2}/4-m_u^2(T)$, as above. Thus, we see that the phase
space factor has an exponent of 1/2 corresponding to a
\textit{s}-wave amplitude. For the scalars, the exponent of the
phase-space factor is 3/2, as seen in Eq.\,(A6).

For a study of vector mesons we consider \be
\beta_{\mu\nu}^V=\mbox{Tr}[\,\gamma_\mu(\slr k+m_1)\gamma_\nu(\slr
k-\slr P+m_2)\,]\,,\ee and calculate \be
g^{\mu\nu}\beta_{\mu\nu}^V=4[\,P^2-m_1^2-m_2^2+4m_1m_2\,]\,,\ee
which, in the equal-mass case, is equal to $4P_0^2+8m^2(T)$, when
$\vec P=0$. This result is needed when we calculate the correlator
of vector currents. Note that, for the elevated temperatures
considered in this work, $m_u(T)=m_d(T)$ is quite small, so that
$4P_0^2+8m_u^2(T)$ can be approximated by $4P_0^2$, when we
consider the vector current correlation functions. In that case,
we have \be \mbox{Im}\,J_V(P^2,T) \simeq
\frac{2}{3}\mbox{Im}\,J_P(P^2,T)\,.\ee At this point it is useful
to define functions that do not contain that Gaussian regulator:
\be\mbox{Im}\,\tilde{J}_P(P^2,T)=\frac{N_cP_0^2}{8\pi}\left(1-\frac{4m^2(T)}{P_0^2}\right)^{1/2}[\,1-2n_1(k)\,]\,,\ee
and
\be\mbox{Im}\,\tilde{J}_V(P^2,T)=\frac{2}{3}\frac{N_cP_0^2}{8\pi}\left(1-\frac{4m^2(T)}{P_0^2}\right)^{1/2}[\,1-2n_1(k)\,]\,,\ee
For the functions defined in Eq.\,(A14) and (A15) we need to use a
twice-subtracted dispersion relation to obtain
$\mbox{Re}\,\tilde{J}_P(P^2,T)$, or
$\mbox{Re}\,\tilde{J}_V(P^2,T)$. For example,
\be\mbox{Re}\,\tilde{J}_P(P^2,T)=\mbox{Re}\,\tilde{J}_P(0,T)+
\frac{P^2}{P_0^2}[\,\mbox{Re}\,\tilde{J}_P(P_0^2,T)-\mbox{Re}\,\tilde{J}_P(0,T)\,]\\\nonumber
+\frac{P^2(P^2-P_0^2)}{\pi}\int_{4m^2(T)}^{\tilde{\Lambda}^{2}}
ds\frac{\mbox{Im}\,\tilde{J}_P(s,T)}{s(P^2-s)(P_0^2-s)}\,,\ee
where $\tilde{\Lambda}^{2}$ can be quite large, since the integral
over the imaginary part of the polarization function is now
convergent. We may introduce $\tilde{J}_P(P^2,T)$ and
$\tilde{J}_V(P^2,T)$ as complex functions, since we now have both
the real and imaginary parts of these functions. We note that the
construction of either $\mbox{Re}\,J_P(P^2,T)$, or
$\mbox{Re}\,J_V(P^2,T)$, by means of a dispersion relation does
not require a subtraction. We use these functions to define the
complex functions $J_P(P^2,T)$ and $J_V(P^2,T)$.

In order to make use of Eq.\,(A16), we need to specify
$\tilde{J}_P(0)$ and $\tilde{J}_P(P_0^2)$. We found it useful to
take $P_0^2=-1.0$ \gev2 and to put $\tilde{J}_P(0)=J_P(0)$ and
$\tilde{J}_P(P_0^2)=J_P(P_0^2)$. The quantities $\tilde{J}_V(0)$
and $\tilde{J}_V(P_0^2)$ are determined in an analogous function.
This procedure in which we fix the behavior of a function such as
$\mbox{Re}\tilde{J}_V(P^2)$ or $\mbox{Re}\tilde{J}_V(P^2)$ is
quite analogous to the procedure used in Ref.\,[\,20\,]. In that
work we made use of dispersion relations to construct a continuous
vector-isovector current correlation function which had the
correct perturbative behavior for large $P^2\rightarrow-\infty$
and also described the low-energy resonance present in the
correlator due to the excitation of the $\rho$ meson. In
Ref.\,[\,20\,] the NJL model was shown to provide a quite
satisfactory description of the low-energy resonant behavior of
the vector-isovector correlation function.

We now consider the calculation of temperature-dependent hadronic
current correlation functions. The general form of the correlator
is a transform of a time-ordered product of currents, \be iC(P^2,
T)=\int d^4xe^{iP\cdot x}<\!\!<T(j(x)j(0))>\!\!>\,,\ee where the
double bracket is a reminder that we are considering the finite
temperature case.

For the study of pseudoscalar states, we may consider currents of
the form $j_{P,i}(x)=\tilde{q}(x)i\gamma_5\lambda^iq(x)$, where,
in the case of the $\pi$ mesons, $i=1,2$ and $3$. For the study of
scalar-isoscalar mesons, we introduce
$j_{S,i}(x)=\tilde{q}(x)\lambda^i q(x)$, where $i=0$ for the
flavor-singlet current and $i=8$ for the flavor-octet current.

In the case of the pseudoscalar-isovector mesons, the correlator
may be expressed in terms of the basic vacuum polarization
function of the NJL model, $J_P(P^2, T)$. Thus, \be C_P(P^2,
T)=J_P(P^2, T)\frac{1}{1-G_{P}(T)J_P(P^2, T)}\,,\ee where $G_P(T)$
is the coupling constant appropriate for our study of $\pi$
mesons. We have found $G_P(T)=13.49$\gev{-2} by fitting the pion
mass in a calculation made at $T=0$, with $m_u = m_d =0.364$ GeV.
The result given in Eq.\,(A18) is only expected to be useful for
small $P^2$, since the Gaussian regulator strongly modifies the
large $P^2$ behavior. Therefore, we suggest that the following
form is useful, if we are to consider the larger values of $P^2$.
\be \frac{C_{P}(P^2, T)}{P^2}=\left[\frac{\tilde{J}_P(P^2,
T)}{P^2}\right] \frac{1}{1-G_P(T)J_P(P^2, T)}\,.\ee (As usual, we
put $\vec{P}=0$.) This form has two important features. At large
$P_0^2$, ${\mbox{Im}\,C_{P}(P_0, T)}/{P_0^2}$ is a constant, since
${\mbox{Im}\,\tilde{J}_{P}(P_0^2, T)}$ is proportional to $P_0^2$.
Further, the denominator of Eq.\,(A19) goes to 1 for large
$P_0^2$. On the other hand, at small $P_0^2$, the denominator is
capable of describing resonant enhancement of the correlation
function. As we have seen, the results obtained when Eq.\,(A19) is
used appear quite satisfactory. (\,We may again refer to
Ref.\,[\,20\,], in which a similar approximation is described.)

For a study of the vector-isovector correlators, we introduce
conserved vector currents $j_{\mu,
i}(x)=\tilde{q}(x)\gamma_{\mu}\lambda_i q(x)$ with i=1, 2 and 3.
In this case we define \be J_V^{\mu\nu}(P^2,
T)=\left(g\,{}^{\mu\nu}-\frac{P\,{}^\mu
P\,{}^\nu}{P^2}\right)J_V(P^2, T)\ee and \be C_V^{\mu\nu}(P^2,
T)=\left(g\,{}^{\mu\nu}-\frac{P\,{}^\mu
P\,{}^\nu}{P^2}\right)C_V(P^2, T)\,,\ee taking into account the
fact that the current $j_{\mu,\,i}(x)$ is conserved. We may then
use the fact that \be J_V(P^2,T) =
\frac13g_{\mu\nu}J_V^{\mu\nu}(P^2,T)\ee and
\be\mbox{Im}\,J_V(P^2,T)&=&
\frac23\left[\frac{P_0^2+2m_u^2(T)}{8\pi}\right]
\left(1-\frac{4m_u^2(T)}{P_0^2}\right)^{1/2}e^{-\vec
k\,{}^2/\alpha^2}[\,1-2n_1(k)\,]\\
&\simeq& \frac{2}{3}\mbox{Im}J_P(P^2,T)\,.\ee (See Eq.\,(A7) for
the specification of $k=|\vec k|$.) We then have \be
C_V(P^2,T)=\tilde{J}_V(P^2,T)\frac1{1-G_V(T)J_V(P^2,T)}\,,\ee
where we have introduced \be\mbox{Im}\tilde{J}_V(P^2,T)&=&
\frac23\left[\frac{P_0^2+2m_u^2(T)}{8\pi}\right]
\left(1-\frac{4m_u^2(T)}{P_0^2}\right)^{1/2}[\,1-2n_1(k)\,]\\
&\simeq& \frac{2}{3}\mbox{Im}\tilde{J}_P(P^2,T)\,. \ee
In the literature, $\omega$ is used instead
of $P_0$ [\,4-6\,]. We may define the spectral functions \be\sigma_V(\omega,
T)=\frac{1}{\pi}\,\mbox{Im}\,C_V(\omega, T)\,,\ee and \be\sigma_P(\omega,
T)=\frac{1}{\pi}\,\mbox{Im}\,C_P(\omega, T)\,,\ee

Since different conventions are used in the literature [\,4-6\,],
we may use the notation $\overline{\sigma}_P(\omega, T)$ and
$\overline{\sigma}_V(\omega, T)$ for the spectral functions given
there. We have the following relations: \be
\overline{\sigma}_P(\omega, T)=\sigma_P(\omega, T)\,,\ee and
\be\frac{\overline{\sigma}_V(\omega,
T)}{2}=\frac{3}{4}\sigma_V(\omega, T)\,,\ee where the factor 3/4
arises because, in Refs. [\,4-6\,], there is a division by 4,
while we have divided by 3, as in Eq.\,(A22).

\section{}
\renewcommand{\theequation}{B\arabic{equation}}

Here we extend the work of Appendix A to consider case of finite
three-momentum, $\vec{P}$. We consider the calculation of
$\mbox{Im}J_P(P^0,\vec{P},T)$. The momenta $P^0$ and $\vec{P}$ are
the values external to the loop diagram. Internal to the diagram,
we have a quark of momentum $k^\mu+P^\mu/2$ leaving the left-hand
vertex and an antiquark of momentum $k^\mu-P^\mu/2$ entering the
left-hand vertex. It is useful to define \be
E_1(k)&=&\left|\vec{k}+\vec{P}/2\right|\\
&=&\left(k^2+\frac{P^2}4+kP\cos\theta\right)^{1/2}\ee and
\be E_2(k)&=&\left|\vec{k}-\vec{P}/2\right|\\
&=&\left(k^2+\frac{P^2}4-kP\cos\theta\right)^{1/2}\,.\ee Here
$k=|\vec{k}|$ and $P=|\vec{P}|$.

We have \be \mbox{Im}\,J_P(P^0,\vec{P},
T)=\frac12N_c\beta_P\,\epsilon(P^0)\mytint ke^{-\vec
k\,{}^2/\alpha^2}\left(\frac{2\pi}{2E_1(k)2E_2(k)}\right)\\\nonumber
\times\{[1-n_1(k)-n_2(k)]
\delta(P\,{}^0-E_1(k)-E_2(k))\\\nonumber-[n_1(k)-n_2(k)]
\delta(P\,{}^0+E_1(k)-E_2(k))\\\nonumber-[n_2(k)-n_1(k)]
\delta(P\,{}^0-E_1(k)+E_2(k))\\\nonumber-[1-n_1(k)-n_2(k)]
\delta(P\,{}^0+E_1(k)+E_2(k))\}\,.\ee Here, \be
n_1(k)=\frac1{e^{\,\beta E_1(k)}+1}\,,\ee and \be
n_2(k)=\frac1{e^{\,\beta E_2(k)}+1}\,.\ee In Eq. (B5), the second
and third terms cancel and the fourth term does not contribute. It
is useful to rewrite $\delta(P^0-E_1(k)-E_2(k))$ using \be
\delta[f(\cos\theta)]=\frac2{\left|\frac{\partial f}{\partial \cos
\theta}\right|_x}\delta(\cos\theta-x)\,,\ee where \be
x^2&=&\cos^2\theta\\&=&\frac{4P_0^2(k^2+P^2/4)-P_0^4}{4k^2P^2}\nonumber\,.\ee
We find \be \left|\frac{\partial f}{\partial \cos
\theta}\right|=\frac12kP\left|\frac{E_1(k)-E_2(k)}{E_1(k)E_2(k)}\right|\,,\ee
and obtain \be \mbox{Im}\,J_P(P^0,\vec{P},
T)=\frac12N_c\beta_P\,\epsilon(P^0)(2\pi)^2\int
\frac{k^2dk}{(2\pi)^3}e^{-k\,{}^2/\alpha^2}\\\nonumber
\int\frac1{2E_1(k)E_2(k)}[1-n_1(k)-n_2(k)]
\left|\frac{\partial{f(\cos\theta)}}{\partial{
\cos\theta}}\right|\\\nonumber
\times\delta(\cos\theta-x)d(\cos\theta)\,.\ee We note there is a
singularity when $E_1(k)=E_2(k)$. That occurs when $\cos\theta=0$
or $\theta=\pi/2$. For our calculations we eliminate the point
with $\theta=\pi/2$ when evaluating the angular integral over
$d(\cos\theta)\delta(\cos\theta-x)$ in the last expression. We
obtain \be \mbox{Im}\,J_P(P^0,\vec{P},
T)=N_c\beta_P\,\epsilon(P^0)\frac{4\pi^2}{(2\pi)^3}\int^{k_{max}}
k^2dk\,e^{-k\,{}^2/\alpha^2}\\\nonumber
\times\left.\frac{[1-n_1(k)-n_2(k)]}{kP|E_1(k)-E_2(k)|}\right|_x\,,\ee
where \emph{x} is obtained from Eq. (B9), \be
x=\frac{P^0}{kP}\left[k^2+\frac{P^2}4-\frac{P_0^2}4\right]^{1/2}\ee

For the calculations reported in this work we have
$P^{0}=E(\overrightarrow{p_{1}})+E(\overrightarrow{p_{2}})$ and
$\overrightarrow{P}=\overrightarrow{p_{1}}+\overrightarrow{p_{2}}$,
where $\overrightarrow{p_{2}}$ is the quark momentum and
$\overrightarrow{p_{1}}$ is the antiquark momentum. Thus, we may
also use the notation
$J(\overrightarrow{p_{1}},\overrightarrow{p_{2}})$ as we have done
in the main text.



\end{document}